# Grandma's Local Realistic Theory for the Einstein-Podolsky-Rosen-Bohm Experiment

Boon Leong Lan

*Sch. of Eng. & Sci., Monash University, 2 Jalan Kolej, 46150 PJ, Selangor, Malaysia*

lan.boon.leong@engsci.monash.edu.my

A local realistic theory is presented for Mermin's special case of the EPRB experiment. The theory, which is readily extended to the general EPRB experiment, reproduces all the predictions of quantum theory. It also reveals that Bell, and also Hess and Philipp, had made an error in the mathematical formulation of Einstein's locality or no-action-at-a-distance principle.

## I. INTRODUCTION

In the Einstein-Podolsky-Rosen-Bohm (EPRB) experiment, a source produces a system of two spin-1/2 particles (labeled *1* and *2*) that fly apart in opposite directions, each towards a Stern-Gerlach magnet: particle *1* (*2*) towards magnet *1* (*2*). Each magnet can be rotated in a plane perpendicular to the line of flight of the particles; the angle $\theta$ ($\phi$) gives the direction of magnet *1* (*2*). More than twenty years ago, Mermin [1] considered a special case of the EPRB experiment where each magnet has only three possible angle settings (labeled *a*, *b*, and *c*) and presented a local realistic theory for the experiment. If the two magnet angles are the same, Mermin's local realistic theory predicts [1] that particle *1* and particle *2* will be measured to have opposite spins with probability one, in agreement with the prediction of quantum theory. On the other hand, if the two magnet angles are different, Mermin's local realistic theory predicts [1] that the probability of measuring opposite spins is at least one-third. However, this is not always the prediction of quantum theory: for instance [1], if the two magnet angles differ by $\pm 120^{\circ}$ or $\pm 240^{\circ}$, the probability of measuring opposite spins is one-fourth. In his original paper [1], and also in a later more refined version [2], Mermin challenged the reader to construct a local realistic theory for his special case of the EPRB experiment that will yield the same predictions as quantum theory in all cases, i.e., in the case where the magnet angles are the same and in the case where the magnet angles are different. Such a theory, which is readily extended to the general case where each magnet can be set at any angle, is presented in this paper. I have chosen to call it Grandma's local realistic theory.

The rest of the paper is organized as follows. Quantum theory for the EPRB experiment is summarized in Sec. II. Mermin's local realistic theory is recapped in Sec. III. Grandma's local realistic theory, which is based on a preliminary work published in the proceedings [3] of the 2001 Garda Workshop, is presented and discussed in Sec. IV.

**II. QUANTUM THEORY**

Let $|\pm,\theta\rangle$ ($|\pm,\phi\rangle$) be the eigenstates of the projection of the spin operator of particle *1* (*2*) onto the unit vector in the direction of magnet *1* (*2*). The spin part of the quantum wave function for the system of two spin-1/2 particles, known as the singlet state in the literature, can be expressed [4] in terms of the four product states $\{|\pm,\theta\rangle|\pm,\phi\rangle\}$ that I call system spin states:

$$|S\rangle = \frac{1}{\sqrt{2}}\left[-i\sin\left(\frac{\theta-\phi}{2}\right)|+,\theta\rangle|+,\phi\rangle + \cos\left(\frac{\theta-\phi}{2}\right)|+,\theta\rangle|-,\phi\rangle\right.$$

$$\left.-\cos\left(\frac{\theta-\phi}{2}\right)|-,\theta\rangle|+,\phi\rangle + i\sin\left(\frac{\theta-\phi}{2}\right)|-,\theta\rangle|-,\phi\rangle\right] \quad (1)$$

The quantum conditional probabilities for the possible joint measurement outcomes are easily determined from the expansion coefficients in Eq. (1):

$$P(++|\theta,\phi) = P(--|\theta,\phi) = \frac{1}{2}\sin^2\left(\frac{\theta-\phi}{2}\right), \quad (2)$$

$$P(+-|\theta,\phi) = P(-+|\theta,\phi) = \frac{1}{2}\cos^2\left(\frac{\theta-\phi}{2}\right), \quad (3)$$

where, for instance, $P(+-|\theta,\phi)$ is the probability of measuring spin up for particle *1* and spin down for particle *2* given that magnet *1* is set at angle $\theta$ and magnet *2* is set at angle $\phi$.

## III. MERMIN'S LOCAL REALISTIC THEORY

In an experimental run, prior to measurement, particle *1* is actually in one of its two possible spin states, $|+,\theta\rangle$ or $|-,\theta\rangle$, for each of the three possible settings (*a*, *b*, *c*) of the angle $\theta$ of magnet *1*. Particle *2* is also actually in one of its two possible spin states, $|+,\phi\rangle$ or $|-,\phi\rangle$, for each of the three possible settings (*a*, *b*, *c*) of the angle $\phi$ of magnet *2*. For magnet angles $\theta$ and $\phi$ which are the same, the pre-existing spin states of particle *1* and particle *2* are opposite to each other, i.e., if particle *1* is spin up, particle *2* is spin down, and vice-versa. For example, if $|+,\theta=a\rangle$, $|+,\theta=b\rangle$, and $|-,\theta=c\rangle$ are the pre-existing states of particle *1*, then $|-,\phi=a\rangle$, $|-,\phi=b\rangle$, and $|+,\phi=c\rangle$ are the pre-existing states of particle *2*.

In an experimental run, for a chosen pair of magnet angles, measurement reveals the pre-existing spin state of particle *1* for the chosen magnet angle $\theta$ and the pre-existing spin state of particle *2* for the chosen magnet angle $\phi$. For the pre-existing states in the example above, if the chosen magnet angles are $\theta = a$ and $\phi = c$, then both particles will be measured spin up. If the chosen magnet angles are $\theta = a$ and $\phi = a$, then the two particles will be measured to have opposite spins: in particular, spin up for particle *1* and spin down for particle *2*. In general, regardless of the pre-existing spin states of the two particles, measurement will always yield opposite spins for the two particles if the two chosen magnet angles are the same.

The pre-existing spin states of particle *1* (*2*) depend on the possible angles of magnet *1* (*2*) but not on the possible angles of magnet *2* (*1*). Therefore, the measurement result *A* (*B*) for particle *1* (*2*) depends on the chosen angle $\theta$ ($\phi$) for magnet *1* (*2*) but not on the chosen angle $\phi$ ($\theta$) for magnet *2* (*1*):

$A(\theta)$ and $B(\phi)$.     (4)

As an illustration, in the example above, if $\theta = a$ is the chosen angle of magnet *1*, particle *1* will measured spin up regardless of the angle $\phi$ chosen for magnet *2*.

From an orthodox point of view, in the EPRB experiment, each particle does not have a definite spin prior to observation. Measurement of one particle compels that particle to acquire a definite spin and instantaneously triggers the other particle, spatially separated from the first, to also acquire a definite spin. This instantaneous 'triggering' is non-locality or action at a distance [2,5], which Einstein considered spooky and unacceptable. Bell [6], in the derivation of his famous inequality, assumed that locality or no action at a distance requires that the measurement result *A* (*B*) for particle *1* (*2*) does not depend on the chosen angle $\phi$ ($\theta$) for magnet *2* (*1*). Mermin [1,2] incorporated Bell's requirement for locality in his realistic theory above (see previous paragraph).

## IV. GRANDMA'S LOCAL REALISTIC THEORY

In an experimental run, prior to measurement, the system of two spin-1/2 particles is actually in one of the four system spin states, $|+,\theta\rangle|+,\phi\rangle$ or $|-,\theta\rangle|-,\phi\rangle$ or $|+,\theta\rangle|-,\phi\rangle$ or $|-,\theta\rangle|+,\phi\rangle$, for each of the nine possible pairs of magnet angles, i.e., the system is actually in nine system spin states, one system spin state for each possible pair of magnet angles. For each possible pair of magnet angles, $\theta$ and $\phi$, the system is actually either in state $|+,\theta\rangle|+,\phi\rangle$ with probability $(1/2)\sin^2\delta$, or in state $|-,\theta\rangle|-,\phi\rangle$ with probability $(1/2)\sin^2\delta$, or in state $|+,\theta\rangle|-,\phi\rangle$ with probability $(1/2)\cos^2\delta$, or in state $|-,\theta\rangle|+,\phi\rangle$ with probability $(1/2)\cos^2\delta$, where $\delta = (\theta-\phi)/2$. So, for magnet angles $\theta$ and $\phi$ which are the same, the system is definitely not in state $|+,\theta\rangle|+,\phi\rangle$ or in state $|-,\theta\rangle|-,\phi\rangle$; it is actually either in state $|+,\theta\rangle|-,\phi\rangle$ or state $|-,\theta\rangle|+,\phi\rangle$ with equal probability of 1/2. For magnet angles $\theta$ and $\phi$ which differ by $\pm 180°$, the system is definitely not in state $|+,\theta\rangle|-,\phi\rangle$ or in state $|-,\theta\rangle|+,\phi\rangle$; it is actually either in state $|+,\theta\rangle|+,\phi\rangle$ or state $|-,\theta\rangle|-,\phi\rangle$ with equal probability of 1/2. Table 1 lists a possible set of nine *pre-existing* system spin states, one state for each *possible* pair of magnet angles, that pre-existed before *any* pair of magnet angles is *chosen* in a hypothetical experimental run.

In an experimental run, for the chosen pair of magnet angles, measurement reveals the pre-existing system spin state for that pair of angles. For example, suppose, in an experimental run, the set of nine pre-existing system spin states is given by Table 1. If the chosen magnet angles are $\theta = a$ and $\phi = a$, then measurement will yield spin up for particle *1* and spin down for particle *2*, revealing that, for this pair of magnet angles, the pre-existing system spin state is $|+, \theta = a\rangle |-, \phi = a\rangle$. If the chosen magnet angles are $\theta = a$ and $\phi = b$, then the pre-existing state $|-, \theta = a\rangle |-, \phi = b\rangle$ will be revealed by measurement. Similarly, if the chosen pair of magnet angles is one of the other seven possibilities, the corresponding pre-existing system spin state listed in Table 1 will be revealed by measurement.

**Table 1.** *Pre-existing* system spin states, one state for each *possible* pair of magnet angles, in a hypothetical experimental run.

| Possible Angle $\theta$ For Magnet *1* | Possible Angle $\phi$ For Magnet *2* | Pre-existing System Spin State |
|:---:|:---:|:---:|
| a | a | $|+, \theta = a\rangle |-, \phi = a\rangle$ |
| a | b | $|-, \theta = a\rangle |-, \phi = b\rangle$ |
| a | c | $|-, \theta = a\rangle |-, \phi = c\rangle$ |
| b | a | $|-, \theta = b\rangle |-, \phi = a\rangle$ |
| b | b | $|+, \theta = b\rangle |-, \phi = b\rangle$ |
| b | c | $|+, \theta = b\rangle |+, \phi = c\rangle$ |
| c | a | $|+, \theta = c\rangle |+, \phi = a\rangle$ |
| c | b | $|+, \theta = c\rangle |+, \phi = b\rangle$ |
| c | c | $|-, \theta = c\rangle |+, \phi = c\rangle$ |

Because measurement merely reveals the *pre-existing* system spin state for the chosen pair of magnet angles, there isn't any action at a distance whatsoever. In other words, Grandma's realistic theory is intrinsically local.

In an experimental run, for the chosen pair of magnet angles $\theta$ and $\phi$, since the probability that the system is actually in state $|+,\theta\rangle|+,\phi\rangle$ ($|-,\theta\rangle|-,\phi\rangle$) is $(1/2)\sin^2\delta$ where $\delta = (\theta-\phi)/2$, the probability of measuring spin up (down) for particle *1* and spin up (down) for particle *2* is also $(1/2)\sin^2\delta$, in agreement with the predictions of quantum theory [see Eq. (2)]. Likewise, since the probability that the system is actually in state $|+,\theta\rangle|-,\phi\rangle$ ($|-,\theta\rangle|+,\phi\rangle$) is $(1/2)\cos^2\delta$ where $\delta = (\theta-\phi)/2$, the probability of measuring spin up (down) for particle *1* and spin down (up) for particle *2* is also $(1/2)\cos^2\delta$, in agreement with the predictions of quantum theory [see Eq. (3)]. Thus, Grandma's local realistic theory reproduces all the predictions of quantum theory.

In Grandma's local realistic theory, each particle has nine *pre-existing* spin states, one for each *possible* pair of magnet angles, which pre-existed before *any* pair of magnet angles is *chosen*. Therefore the measurement result (i.e., which of the nine pre-existing spin states is revealed) *A* (*B*) for particle *1* (*2*) depends on the chosen angle $\theta$ for magnet *1* and the chosen angle $\phi$ for magnet *2*:

$$A(\theta,\phi) \text{ and } B(\theta,\phi). \tag{5}$$

As an illustration, consider again the set of pre-existing system spin states in Table 1. Suppose $\theta = a$ is the chosen angle for magnet *1*. If $\phi = a$ is the chosen angle for magnet *2*, then particle *1* will be measured spin up. But if $\phi = b$ is the chosen angle for magnet *2*, then particle *1* will be measured spin down. Now, instead, suppose $\phi = a$ is the chosen angle for magnet *2*. If $\theta = a$ is the chosen angle for magnet *1*, then particle *2* will be measured spin down. But if $\theta = c$ is the chosen angle for magnet *1*, then particle *2* will be measured spin up.

The dependence of the measurement result for each particle on the chosen pair of magnet angles in Eq. (5) **does not** however imply non-locality or action at a distance because the chosen pair of magnet angles merely reveals the corresponding *pre-existing* spin state of each particle. This means that locality or no action at a distance **does not** require the measurement result *A* (*B*) for particle *1* (*2*) not to depend on the chosen angle $\phi$ ($\theta$) for magnet *2* (*1*), contrary to what Bell [6] and his followers, Mermin among them, had assumed. In other words, Bell had made a mistake in his mathematical formulation of Einstein's locality or no-action-at-a-distance principle. The much-debated local realistic theory of Hess and Philipp [7-11] also contains the same mis-formulation of locality as Bell but it is able to reproduce the quantum mechanical expectation value of the product *AB* of the measurement results. This reproduction is possible because, in Hess and Philipp's theory, the measurement result of particle *1* (*2*) is also pre-determined by a hidden variable in magnet *1* (*2*), which depends on the angle setting of the magnet and the time of measurement of the particle, in addition to the hidden variable in the source considered by Bell. In contrast, in Grandma's local realistic theory, the measurement results of both particles for each possible pair of magnet angles are completely pre-determined by the hidden variable in the source.

A local realistic theory has also been constructed [12] for the Greenberger-Horne-Zeilinger experiment [13] that reproduces all the predictions of quantum theory. The theory shows that GHZ [13] had also mis-formulated Einstein's locality or no-action-at-a-distance principle in their local realistic theory for the experiment in the same way that Bell had mis-formulated the principle in his local realistic theory for the EPRB experiment.